\documentstyle[referee,epsf,mnras_cite]{mn}

\input{epsf}

\newcommand{\ba}{\begin{eqnarray}}
\newcommand{\ea}{\end{eqnarray}}
\newcommand{\be}{\begin{equation}}
\newcommand{\ee}{\end{equation}}

\newcommand{\ls}{\mathrel{\raise1.16pt\hbox{$<$}\kern-7.0pt %  <
\lower3.06pt\hbox{{$\scriptstyle \sim$}}}}         %  ~
\newcommand{\gs}{\mathrel{\raise1.16pt\hbox{$>$}\kern-7.0pt %  >
\lower3.06pt\hbox{{$\scriptstyle \sim$}}}}         %  ~

\long\def\comment#1{}

\def\fun#1#2{\lower3.6pt\vbox{\baselineskip0pt\lineskip.9pt
  \ialign{$\mathsurround=0pt#1\hfil##\hfil$\crcr#2\crcr\sim\crcr}}}

\def\be{\begin{equation}}
\def\bea{\begin{eqnarray}}
\def\beaa{\begin{eqnarray*}} 
\def\ee{\end{equation}}  
\def\eea{\end{eqnarray}}
\def\eeaa{\end{eqnarray*}} 
\def\la{\bigl\langle} \def\ra{\bigr\rangle}

\def\dO{d\Omega}

\title{Cosmic microwave background bispectrum 
       and slow roll inflation} 
\author[Alejandro Gangui and J\'er\^ome Martin]
{Alejandro Gangui and J\'er\^ome Martin \\
D\'epartement d'Astrophysique Relativiste et de Cosmologie,\\
Observatoire de Paris--Meudon, CNRS UMR 8629, 92195 Meudon Cedex, France}

\date{Submitted 1999 July 24}

\begin{document}
\maketitle

\begin{abstract}
Recent tentative findings of non-Gaussian structure in the
COBE-DMR dataset have triggered renewed attention to candidate 
models from which such intrinsic signature could arise. In the 
framework of slow roll inflation with built-in non
linearities in the inflaton field evolution we present expressions
for both the cosmic microwave background (CMB) skewness and the 
full angular bispectrum ${\cal C}_{\ell_1 \ell_2 \ell_3}$ in terms 
of the slow roll parameters. 
We use an estimator for the angular bispectrum recently proposed 
in the literature and calculate its variance for an arbitrary $\ell_i$
multipole combination. We stress that a real detection of non-Gaussianity 
in the CMB would imply that an important component of the anisotropies
arises from processes {\it other} than primordial quantum fluctuations. 
We further investigate the behavior of the signal-to-(theoretical) 
noise ratio and demonstrate for generic inflationary models 
that it decreases in the limited range of small-$\ell$'s
considered for increasing multipole $\ell$
while the opposite applies for the standard ${\cal C}_{\ell}$'s. 
\end{abstract}


\section{Introduction}

The theory of inflation provides an elegant means to solve the usual
problems (horizon and flatness problems) of the standard model of
Cosmology \cite{Guth}. It consists in assuming that a phase of
accelerated expansion took place in the very early Universe at the GUT
energy scale. In the most simple models of inflation, this phase of
accelerated expansion is driven by a scalar field. The physical origin
of this field is still an open problem and Physics at GUT scale could
be much more complicated than assumed in these simple models. However,
it should be stressed that the very concept of inflation lies in the
fact that the second cosmic time derivative of the scale factor was
positive in the early Universe and is, in this sense, independent of
any model-building provided that the effective inflaton potential is
flat enough to allow an inflationary expansion of at least $\approx
70 \mbox{e-folds}$. Therefore {\it slow roll} inflation can be viewed as a
generic framework which permits to implement concretely the concept of
inflation and allows to perform simple analytical calculations.

The beauty of the inflationary scenario is that, combined with Quantum
Mechanics, it also provides a natural explanation of the origin of the
large scale structures and of the cosmic microwave background (CMB)
radiation anisotropies observed in our Universe
\cite{GuPi,Staro,Haw,BST}. In this explanation, the quantum character
of the inflaton field plays a crucial role since the seeds of these
perturbations are the unavoidable quantum fluctuations present at the
beginning of the inflationary epoch. Then these fluctuations are
parametrically amplified during the accelerated phase of expansion
\cite{Gri}. Therefore the properties of the initial spectrum of
perturbations depend on the initial quantum state in which the
fluctuations were placed and on the behaviour of the scale factor
during inflation. Observationally, we have access to the initial
spectrum when one looks at the CMB anisotropy multipole moments
corresponding to the largest angular scales on the celestial
sphere. Indeed these multipoles are dominated by modes whose
wavelengths are comparable to the size of the horizon today. This
means that after their creation these modes spent most of their time
outside the Hubble radius and as a consequence were not contaminated
by astrophysical processes: in a certain sense, they can be viewed as
a pure relic of the very early Universe.

Among the many features of the perturbations, the statistical 
properties are certainly of a big importance. In the theory of 
cosmological perturbations of quantum mechanical origin, it is assumed 
that the initial state is the vacuum. This seems to be the most 
natural choice although it was already noticed that it could be 
difficult to understand why the fluctuations of a field which is initially 
out of equilibrium would be placed in this state \cite{Unruh}. It was 
recently shown \cite{MRS} that if one tries to start the evolution 
from a non vacuum initial state, then observations require that 
this state be close to the vacuum. This seems to indicate that 
the vacuum is indeed a reasonable choice. Since each mode of the 
perturbations can be viewed as an oscillator, one immediately reaches 
the conclusion that the corresponding statistical properties must 
be Gaussian (recall that the ground-state wavefunction of an harmonic 
oscillator is a Gaussian function). This constitutes an important 
and generic prediction of inflation.

Another source of cosmological relevant density inhomogeneities arises
in models with topological defects of the vacuum, like cosmic strings
and textures. These would leave different imprints on the CMB both at
recombination and later, during the photon travel from the surface of
last scattering to the present, on various angular scales
\cite{alle,mague,battye,conta,pova99,dgs96,ulpen}.  There is the hope
that future balloon-borne (e.g., MAXIMA, \pcite{balloon1}; BOOMERanG,
\pcite{balloon2}) and satellite 
(MAP and Planck surveyor, \pcite{satellite2}) missions
will allow a clean distinction among these different classes of models
by mapping the CMB with unprecedent precision.

A number of authors (e.g.,
\pcite{smoot94,torres95,hinshaw94,hinshaw95,kogut96}) have analysed
the COBE-DMR sky maps with a variety of test (like three-point
statistics, genus, and extrema correlation function) and found perfect
agreement with a Gaussian distribution. Recently however, three groups
\cite{ferre98,pando98,bromley99} have analysed the COBE-DMR four-year
dataset and reported detections of non-Gaussianity casting doubts on
these early findings. Banday {\rm et~al.} 1999 have further analyzed
the same data, finding that the non-Gaussian signal is driven by the
53 GHz sky maps. They concluded that this frequency dependence strongly
indicates that the signal is not of primordial origin.
%Besides the fact of not knowing whether these results are really due
%to primordial fluctuations or to foregrounds/artefacts in the data, 
Despite all this, it seems now that due to limited signal-to-noise, sky
coverage, and uncertainty in foreground substraction, present day
experiments cannot conclusively exclude non-Gaussianity to a
satisfactory confidence level.

The above remarks regarding the Gaussian character of the primordial
perturbations have been established within the framework of the linear
theory of cosmological pertubations. It is clear that generic higher
order, in particular quadratic, terms are present and will produce a
non vanishing signal even for inflationary models \cite{LM}. The
predictions for the three-point correlation function on large angular
scales due to nonlinearities in the inflaton evolution were considered
in the past \cite{fa93,ganguietal94,ga94}.

The post recombination Rees-Sciama effect, due to the mildly
non-linear evolution of the perturbations also contributes to the
signal \cite{lu93,molle,mun}.  Evolving networks of topological
defects continuously seed perturbations on the CMB that, by the very
nature of the sources, are predicted to be highly non-Gaussian
\cite{bouchet,avel,ga95,gamo96}.  Further secondary effects contribute
to produce non-Gaussianities at smaller scales \cite{nabila} and would
be characterized by detectable correlations between gravitational
lensing and Sunyaev-Zel'dovich maps \cite{sperg,sperg2}.

We here present a general discussion of non-Gaussian features arising
in the framework of slow roll inflation.
Our article borrows some definitions and formulas for the CMB
three-point correlation function (specially in section 2) from 
(Gangui et al 1994). Our main aim here is to present
explicitly the derivations of the non-Gaussian estimators as a 
function of the inflationary slow roll parameters in the Legendre
space, namely the full angular bispectrum 
${\cal C}_{\ell_1 \ell_2 \ell_3}$.
Then, in the third section, we present for the first time the
analytical expression for the variance of an estimator recently 
proposed for the bispectrum \cite{ferre98} in the mildly non-Gaussian
approximation.  
This allows us, in the last section, to compare the behaviour of both
quantities for various multipoles and conclude with the by now 
established result supporting the view that the recently observed 
non-Gaussianity cannot be explained in the framework 
of slow roll inflation. 

\section{Non-Gaussian signal in real and Legendre spaces}

In the framework of the theory of cosmological perturbations of
quantum mechanical origin, $\Delta T/T (\vec x , \hat\gamma)$ is an
operator. The corresponding statistical properties are then calculated
by ``sandwiching'' this operator (or a combination of these operators)
with the quantum state $|\Psi \rangle $ in which the quantum
perturbations are placed. However, it has been shown in \cite{jerogri}
that one can think to $\Delta T/T (\vec x , \hat\gamma)$ as a
classical stochastic process. This stochastic process can be expanded
in spherical harmonics
\begin{equation}
\label{eq1}
{\Delta T \over T} (\vec x , \hat\gamma) = \sum_{\ell,m}
a_{\ell}^{m}(\vec x)
{\cal W}_{\ell}
Y_{\ell}^{ m} (\hat\gamma),
\end{equation}
where ${\cal W}_l$ represents the window function of the 
particular experiment. The coefficients $a_{\ell}^{m}(\vec x)$ 
are random variables and are in principle different for different
observers at positions $\vec x$. The statistical properties of 
$\Delta T/T (\vec x , \hat\gamma)$ are completely specified if 
the probability density function (pdf) of the $a_{\ell}^{m}(\vec x)$'s is 
known. Choosing the initial state to be the vacuum, i.e. $|\Psi \rangle =
|0 \rangle $, and considering only linear terms is equivalent to
saying that 
the pdf of the $a_{\ell}^{m}(\vec x)$'s is a Gaussian 
distribution \cite{jerogri}. This means that
\begin{equation}
\label{statgau}
\la a_\ell^m (\vec x) \ra = 0, \quad  
\la a_{\ell_1}^{m_1}(\vec x) {a_{\ell_2}^{m_2}}^*(\vec x) 
\ra =  {\cal C}_{\ell_1}\delta_{\ell_1\ell_2} \delta_{m_1 m_2}, 
\end{equation} 
where brackets $\la\cdot\ra$ stands for an average over the ensemble
of possible universes in the sense explained above. The variance is
rotationally invariant, i.e. depends only on $\ell$, signaling
statistical isotropy. For Gaussian fields the previous equations are
sufficient since this kind of fields are completely characterized by
giving their two-point correlation function or, equivalently, their
(angular) power spectrum.

We now take into account the non linearities. This means that the pdf
of the $a_{\ell}^{m}(\vec x)$'s is no longer a Gaussian
distribution. The two first moments will still be given by
Eq. (\ref{statgau}) and the difference will show up at the level of
the third order moment. Predictions from different models usually come
as expressions for the ensemble average $\bigl\langle a_{\ell_1
}^{m_1} a_{\ell_2}^{ m_2} a_{\ell_3}^{ m_3} \bigr\rangle$ which can be
written in full analogy with Eq. (\ref{statgau}) in terms of the
angular bispectrum ${\cal C}_{\ell_1 \ell_2 \ell_3}$ as follows:
\be
\label{bicl}
\bigl\langle a_{\ell_1 }^{m_1} a_{\ell_2}^{ m_2} a_{\ell_3}^{ m_3}
\bigr\rangle = 
\left(^{\ell_1~\,\;\ell_2~\,\;\ell_3}_{m_1~m_2~m_3}\right)
{\cal C}_{\ell_1 \ell_2 \ell_3},
\ee
where now the proportionality factor is a Wigner 3-$j$ symbol.
This is non-zero only if the indices
$\ell_i$, $m_i$ ($i=1,2,3$) fulfill the relations: $\vert \ell_j -
\ell_k \vert \leq \ell_i \leq \vert \ell_j + \ell_k \vert$ 
and $m_1 + m_2 + m_3 = 0$. 
There is an additional ``selection rule'' in this equation that  
arises from demanding that
$\bigl\langle\Delta T (\hat\gamma_1) \Delta T (\hat\gamma_2)
 \Delta T (\hat\gamma_3)\bigr\rangle$
be invariant under spatial inversions. One then obtains
$\bigl\langle a_{\ell_1 }^{m_1} a_{\ell_2}^{ m_2} a_{\ell_3}^{ m_3}
\bigr\rangle = 0$ for $\ell_1 + \ell_2 + \ell_3 = odd$
\cite{luo94}.

For a Gaussian model we clearly have ${\cal C}_{\ell_1 \ell_2
\ell_3}=0$ whereas the non linear evolution of the perturbations will
induce a ${\cal C}_{\ell_1 \ell_2 \ell_3}\neq 0$. Therefore in order
to probe the Gaussian character of the stochastic process $\Delta T/T
(\vec x , \hat\gamma)$, it is certainly convenient to study quantities
related to the third moment. {\em A priori} a large choice is
allowed. Here below we will concentrate on the CMB collapsed three-point 
correlation function, defined in real space.
The skewness will just be the particular case of the collapsed
function at zero lag.

The three--point correlation function for points at
three arbitrary angular separations $\alpha$, $\beta$ and $\gamma$ is
given by the average product of temperature fluctuations in all
possible three directions with those angular separations among them
\cite{ganguietal94}. The collapsed case corresponds to the choice
$\alpha=\beta$ and $\gamma=0$ and reads
\begin{equation}
C_3(\alpha) \equiv \int \frac{d\Omega_{\hat \gamma_1}}{4\pi}
\int \frac{d\Omega_{\hat \gamma_2}}{2\pi}
{\Delta T\over T}  (\hat\gamma_1) {\Delta T \over T}^2 (\hat\gamma_2)
\delta(\hat\gamma_1\cdot\hat\gamma_2 -\cos \alpha).
\label{skew1}
\end{equation}
For $\alpha=0$, we recover the well-known expression for the
skewness, $C_3(0)$. Using the spherical harmonics expansion 
(\ref{eq1}) the last equation can be rewritten as: 
\begin{equation}
\label{coll}
C_3(\alpha) =\frac{1}{4\pi}
\sum_{\ell_1,\ell_2,\ell_3,m_1,m_2,m_3}
P_{\ell_1}(\cos\alpha)
a_{\ell_1 }^{m_1} a_{\ell_2}^{ m_2} a_{\ell_3}^{ m_3}
{\cal W}_{\ell_1} {\cal W}_{\ell_2} {\cal W}_{\ell_3}
\bar {\cal H}_{\ell_1 \ell_2 \ell_3}^{m_1 m_2 m_3} ~,
\end{equation}
where we have defined the coefficients 
$\bar {\cal H}_{\ell_1 \ell_2 \ell_3}^{m_1 m_2 m_3}$ by
\begin{equation}
\label{ana91}
\bar {\cal H}_{\ell_1 \ell_2 \ell_3}^{m_1 m_2 m_3}
\equiv \int
d\Omega_{\hat\gamma} Y_{\ell_1}^{m_1} (\hat\gamma)
Y_{\ell_2}^{m_2}(\hat\gamma) Y_{\ell_3}^{m_3} (\hat\gamma) ~,
\end{equation}
which has a simple expression in terms of Wigner 3-$j$ symbols
\cite{messiah}:
\be 
\label{goodh} 
\bar {\cal H}_{\ell_1\ell_2\ell_3}^{m_1 m_2 m_3} = 
\sqrt{\frac{(2\ell_1 +1)(2\ell_2 +1)(2\ell_3 +1)}{4\pi }}
\left(^{\ell_1~\ell_2~\ell_3}_{0~~0~~0}\right) 
\left(^{\ell_1~\,\;\ell_2~\,\;\ell_3}_{m_1~m_2~m_3}\right) ~. 
\ee 
We see here that the condition $\ell_1 + \ell_2 + \ell_3 = even$ is
enforced by the presence of the first Wigner 3-$j$ symbol.
Substitution of Eq. (\ref{bicl}) together with Eq. (\ref{goodh}) 
into Eq. (\ref{coll}), where we have taken the ensemble average, 
yields for the mean collapsed three-point correlation function
\be
\label{ccc3}
\bigl\langle
C_3(\alpha)
\bigr\rangle =
\sum_{\ell_1,\ell_2,\ell_3}
\sqrt{ 2\ell_1 +1\over 4\pi }
\sqrt{ 2\ell_2 +1\over 4\pi }
\sqrt{ 2\ell_3 +1\over 4\pi }
\left(^{\ell_1~\ell_2~\ell_3}_{0~~0~~0}\right)
{\cal W}_{\ell_1} {\cal W}_{\ell_2} {\cal W}_{\ell_3}
{\cal C}_{\ell_1 \ell_2 \ell_3}
P_{\ell_1}(\cos\alpha).
\ee
We see from this that all terms in the sum satisfying 
$\ell_1 + \ell_2 + \ell_3 = even$ and the triangle inequalities 
{\it but otherwise arbitrary} will contribute to the value of 
the collapsed three-point function and hence also to the skewness. 
In general a complete probe of the three-point function will require 
the knowledge of all the coefficients 
${\cal C}_{\ell_1 \ell_2 \ell_3}$ and not just the 
``diagonal'' ones, ${\cal C}_{\ell \ell \ell}$. 
%We will see below how to express the bispectrum in terms of the 
%inflationary slow roll parameters. 

To estimate the amplitude of the non--Gaussian character of the
fluctuations one usually considers the ``dimensionless'' skewness
${\cal S}_1 \equiv \la C_3(0) \ra / \la C_2 (0)\ra ^{3/2}$.
Alternatively, if we want our results to be independent of the
normalisation, we may also define the ratio ${\cal S}_2 \equiv \la
C_3(0)\ra/\la C_2 (0)\ra^2$. Both quantities will be known once 
the bispectrum has been calculated. Therefore our aim is now to
compute ${\cal C}_{\ell_1 \ell_2 \ell_3}$ in the context of slow roll
inflation, i.e. in terms of the inflationary slow roll parameters.

In the framework of the stochastic approach to inflation
\cite{Starosto,GLM}, the calculations reported in (Gangui et al 1994)
valid for models satisfying slow roll dynamics yield
\be 
\label{aaa}
\la a_{\ell_1}^{m_1} a_{\ell_2}^{m_2} a_{\ell_3}^{m_3} \ra 
= 
{15\over 48\pi} \left[ X^2 - 4 m_{\rm Pl} X' \right]
\bigl[ {\cal C}_{\ell_1} {\cal C}_{\ell_2} + 
{\cal C}_{\ell_2} {\cal C}_{\ell_3} + 
{\cal C}_{\ell_3} {\cal C}_{\ell_1} 
\bigr] \bar {\cal H}_{\ell_1 \ell_2 \ell_3}^{m_1 m_2 m_3} \ , 
\ee 
where in general one requires that the inflaton potential $V(\phi)$ be
a smooth function of its argument, which translates into requiring well
defined values for the steepness of the potential $X \equiv m_{\rm Pl}
V'/ V$ (here $' \equiv {\rm d}/{\rm d}\phi$ and $m_{\rm Pl}$ is Planck
mass) and its derivatives throughout the range of relevant scales
\cite{turner93}. Another way of expressing this result is in terms of
the standard slow roll parameters (\pcite{kinney97}; see also
\pcite{SL,liddle94})
\be
\epsilon \equiv {m_{\rm Pl}^2 \over 16\pi} \Biggl({V' \over V}\Biggr)^2
{\rm ~~and~~}
\eta \equiv {m_{\rm Pl}^2 \over 8\pi} 
    \Biggl[ {V'' \over V} - 
    {1\over 2}\Biggl({V' \over V}\Biggr)^2 \Biggr] .
\ee

For the very large scales we focus on here and for standard chaotic
initial conditions in the inflaton field, both parameters satisfy
$\epsilon,\eta \ll 1$. In terms of these, a comparison of 
Eqs. (\ref{bicl}) and (\ref{aaa}) leads to:
\be
\label{bisr}
{\cal C}_{\ell_1 \ell_2 \ell_3} = 
{5\over 2\sqrt{\pi}} (3\epsilon -2\eta) 
\sqrt{(2\ell_1+1)(2\ell_2+1)(2\ell_3+1)}
\bigl[ {\cal C}_{\ell_1} {\cal C}_{\ell_2} +
       {\cal C}_{\ell_2} {\cal C}_{\ell_3} +
       {\cal C}_{\ell_3} {\cal C}_{\ell_1}
\bigr]
\left(^{\ell_1~\ell_2~\ell_3}_{0~~0~~0}\right).
\ee
This equation shows explicitly the full angular bispectrum arising
from generic inflationary models in terms of the slow roll
parameters.  To be specific in the sequel we will take a potential
$V(\phi)\propto \phi^p$, with $p>1$. Then, after evaluation of the
slow roll parameters at Hubble radius crossing for the relevant very
large scales one has $(\epsilon,\eta) =
(p/(p+200),(p-2)/(p+200))$ corresponding to a scalar spectral index $n
= 1 - (2p+4)/(p+200)$. 
The calculations reported in the following sections will be performed
for a quadratic potential, namely, $p = 2$.

With this result in mind, we can now re-express the three-point
correlation function and skewness.  We first need to evaluate the
multipole moments ${\cal C}_\ell$.  For large scales we have 
$P_\Phi(k) \propto k^{n-4}$ with $n$ corresponding to
the primordial index of density fluctuations (e.g., $n=1$ is the
Harrison-Zel'dovich, scale invariant case)
in which case \cite{be87,flm87}
\be 
\label{cl} 
{\cal C}_\ell = 
{\cal C}_2 
{\Gamma(\ell+n/2-1/2)\Gamma(9/2-n/2)\over\Gamma(\ell+5/2-n/2)
 \Gamma(3/2+n/2)} \equiv 
{\cal C}_2 \, \tilde {\cal C}_\ell , 
\ee 
with ${\cal C}_2$ related to the quadrupole power spectrum 
normalization $Q_{\rm rms-PS} = T_0 (5 {\cal C}_2 / 4 \pi)^{1/2}$. 
Sometimes it turns out to be convenient to factorise the quadrupole 
amplitude out by using $\tilde {\cal C}_\ell$. 
In general, the quadrupole depends on the spectral index $n$. Thus,
normalization analyses of datasets yield the estimate of the pair 
$(n,Q_{\rm rms-PS})$. For example, the maximum likelihood 
analysis of the COBE-DMR dataset performed in  
\cite{Bunn97} yields 
$(n,Q_{\rm rms-PS}) = (1.2, 16.2 \mu K)$ while their best fit 
scale invariant normalization that we will use for the numerics 
in the next section is $Q_{\rm rms-PS} = 18.7 \mu K$ (same
as in \pcite{Gorski96}). 

It is now easy to obtain ${\cal S}_1$ and ${\cal S}_2$ in terms 
of the slow roll parameters. They are given by 
\be 
\label{eq63bis} 
{\cal S}_1 = 
{\sqrt{\pi}\over m_{\rm Pl}^2} \sqrt{3 V \over \epsilon}
(3\epsilon -2\eta)
\Biggl(
{\Gamma (3-n)\Gamma (3/2+n/2)\over\Gamma^2(2-n/2)\Gamma (9/2-n/2)}
\Biggr)^{1/2}
{\cal I}_{3/2}(n), 
\ee 
and 
\be 
\label{eq63} 
{\cal S}_2 = 15 \, (3\epsilon -2\eta) {\cal I}_2(n), 
\ee 
where the normalization dependence (in ${\cal S}_1$) is made explicit
by defining the spectral index-dependent geometrical factor 
\be 
\label{eq64} 
{\cal I}_q(n)\! \equiv \! {{1\over 3} 
  \sum_{\ell_1,\ell_2,\ell_3}(2\ell_1\!+\!1) 
(2\ell_2\!+\!1)(2\ell_3\!+\!1) 
  \bigl[\tilde {\cal C}_{\ell_1} \tilde {\cal C}_{\ell_2}\! +\! 
  \tilde {\cal C}_{\ell_2} \tilde {\cal C}_{\ell_3}\! +\!  \tilde 
  {\cal C}_{\ell_3} \tilde {\cal C}_{\ell_1}\bigr]  
{\cal W}_{\ell_1}\! 
  {\cal W}_{\ell_2}\! {\cal W}_{\ell_3} {\cal 
    F}_{\ell_1 \ell_2 \ell_3} \over \left[\sum_\ell (2l+1) \tilde 
    {\cal C}_\ell {\cal W}_\ell^2 \right]^q}. 
\ee 
The exponent $q$ in the denominator takes values 
$3/2$ and $2$ for ${\cal S}_1$ and ${\cal S}_2$, 
respectively. The coefficients ${\cal F}_{\ell_1 \ell_2 \ell_3} \equiv 
(4\pi)^{-2}\int\dO_{\hat\gamma} \int\dO_{\hat\gamma'} 
P_{\ell_1}(\hat\gamma\cdot\hat\gamma') 
P_{\ell_2}(\hat\gamma\cdot\hat\gamma') 
P_{\ell_3}(\hat\gamma\cdot\hat\gamma')$ 
may be suitably expressed in terms of products of 
factorials of $\ell_1$, $\ell_2$ and $\ell_3$, using 
standard relations for Wigner 3-$j$ symbols: in fact we have 
${\cal F}_{k \ell m} = 
\left(^{k~\ell~m}_{0~0~ 0}\right)^{\!2}$. 
%We see clearly from Eqs. (\ref{eq63bis}) and (\ref{eq63}) that  
%${\cal S}_2$ is {\sl independent} of the normalisation, whereas
%${\cal S}_1$  it is not. 
For the COBE-DMR window function, the numerical factors 
${\cal I}_q(n)$ in Eq. (\ref{eq64}) are of order one for 
all interesting values of the primordial scalar spectral index.
Eqs. (\ref{eq63bis}) to (\ref{eq64}) were already
presented in a different form in (Gangui et al 1994). 
Particular cases of these equations have also been displayed 
in \cite{KK,verde}.

\section{Bispectrum estimator and its variance}

When one particular mechanism for the generation of CMB non--Gaussian
features is specified, it is a direct procedure to compute the
analytical angular bispectrum. One such example was shown in the
previous section in the case of slow roll inflation. However, when
dealing with just one realization of a stochastic process, as is the
case for the CMB, all computed quantities come with theoretical error
bars \cite{sc91,sr93}. 
Even though we can analytically compute mean values, when an
actual observation is made there is a non-vanishing 
probability that it will 
fall within a value $\pm\sigma$ apart from the mean. This problem has
been dubbed ``cosmic variance''. To deal with it, one has to
introduce an estimator $\hat{E}$ of the quantity $e$ we seek, i.e. a
random variable such that $\langle \hat{E}\rangle =e$. In this case
the estimator is said to be unbiased. Then one should compute the
variance of the estimator, $\sigma _{\hat E}$, and try to minimize
it. If it turns out that $\sigma _{\hat E}=0$ then we can find $e$
with the help of one realization only (in fact because each
realization gives $e$). In general, we have $\sigma _{\hat E}\neq
0$, and one can show that this is linked to the fact that a stochastic
process cannot be ergodic on a (celestial) sphere \cite{jerogri}. In
that case $\sigma _{\hat E}$ will express the unavoidable error made
when one estimates the mean of a stochastic process from one
realization.

For the standard angular spectrum ${\cal C}_\ell$ the best unbiased 
estimator is \cite{jerogri,max97} 
\be
\label{estim2}
\hat f_\ell = {1\over 2\ell + 1}\sum_m a_{\ell}^{m} {a_{\ell}^{m}}^*
\ee
and it is easy to check that it is unbiased, 
namely $\la \hat f_\ell \ra = {\cal C}_\ell$.
Its variance, the smallest one amongst all possible estimators 
variances, is given by 
\be
\sigma_{\hat f_\ell} = \sqrt{2\over 2\ell + 1} {\cal C}_\ell .
\ee
It is clear that such an optimal strategy should be followed for the 
bispectrum as well (and for the higher order moments). 
The following expression
\be
\label{estim}
\hat f_{\ell_1 \ell_2 \ell_3} =
\sum_{m_1,m_2,m_3}
\left(^{\ell_1~\,\;\ell_2~\,\;\ell_3}_{m_1~m_2~m_3}\right)
a_{\ell_1 }^{m_1} a_{\ell_2}^{ m_2} a_{\ell_3}^{ m_3}
\ee
is an unbiased estimator of the bispectrum 
${\cal C}_{\ell_1 \ell_2 \ell_3}$ since we easily 
check that $\la \hat f_{\ell_1 \ell_2 \ell_3} \ra = 
{\cal C}_{\ell_1 \ell_2 \ell_3}$. Its variance (squared)
\be
\label{vari}
\sigma^2_{\hat f_{\ell_1 \ell_2 \ell_3}} =
\la\hat f^2_{\ell_1 \ell_2 \ell_3}\ra -
\la\hat f_{\ell_1 \ell_2 \ell_3}\ra^2 
\ee
will give us a first indication of the theoretical 
uncertainties we have to deal
with. However, it should be clear that at this level nothing 
tells us that the one given in Eq. (\ref{estim}) is the {\it best} 
unbiased estimator for the bispectrum. As a consequence, working 
with it might be not the best choice and its variance 
might well be not the smallest one. Finding the best 
estimator for the bispectrum is not a trivial task and is presently 
under investigation \cite{Alje}. Foregrounds, detector noise, 
sample variance in the cut sky are among
the additional issues that need be mastered before claiming a real
non-Gaussian detection.

Recently, similar analyses for the computation of the variance for
the estimator of  
$\bigl\langle a_{\ell_1 }^{m_1} a_{\ell_2}^{ m_2} a_{\ell_3}^{ m_3}
\bigr\rangle$ were presented \cite{luo94,heavens98}. 
If compared with the analysis of \cite{luo94} note that we are
not estimating $a_{\ell_1 }^{m_1} a_{\ell_2}^{ m_2} a_{\ell_3}^{ m_3}$
but the statistically isotropic combination of Eq. (\ref{estim}) and
there the presence of the 3-$j$ symbol makes the whole difference. 
Positive detection of intrinsic non-Gaussianity in the COBE-DMR
four-year dataset was recently suggested 
\cite{ferre98,pando98,bromley99}. 
In particular, Ferreira and collaborators, in the attempt to 
unveil an eventually obscured non-Gaussian 
signal in real space, worked in the Legendre space and 
made use of an estimator in the lines of Eq. (\ref{estim}) above 
but with $\ell_1 = \ell_2 = \ell_3 \equiv \ell$.
They also normalized it by dividing 
$\hat f_{\ell\ell\ell}$ by the estimator of ${\cal C}_{\ell}$ 
[$\hat f_\ell$ in the notation of Eq. (\ref{estim2})] to the power 3/2.

We expect departures from Gaussianity to be weak and hence
neglect the contribution of $\la\hat f_{\ell_1 \ell_2 \ell_3}\ra^2$
to Eq. (\ref{vari}). In this mildly non-Gaussian approximation and
after some straight algebra we obtain
\be
\label{vari2}
\sigma^2_{\hat f_{\ell_1 \ell_2 \ell_3}} =
{\cal C}_{\ell_1} {\cal C}_{\ell_2} {\cal C}_{\ell_3}
(1+\delta_{\ell_1\ell_2}+\delta_{\ell_2\ell_3}+\delta_{\ell_3\ell_1}
+ 2 ~ \delta_{\ell_1\ell_2}\delta_{\ell_2\ell_3}),
\ee
where we demanded $\ell_1 + \ell_2 + \ell_3 = even$ 
(otherwise ${\cal C}_{\ell_1 \ell_2 \ell_3} = 0$) and 
$\ell_i \not= 0$, what considerably simplified the resulting 
expression. We note in passing that for the above computation it is 
useful to recall the identity (Mollerach et al 1995)
\be
\label{foo}
\sum_{m=-\ell}^{\ell}
(-1)^m
\left(^{~\,\;\ell~\,\;\ell~\,\;2k}_{-m~m~~0}\right)
= (-1)^\ell \sqrt{2\ell+1} \; \delta_{k,0} .
\ee
We are now in a position to compare the 
signal, i.e. the bispectrum given in Eq. (\ref{bisr}), with the 
theoretical noise characterized by (\ref{vari2}).

\section{Discussion and conclusions}

In the previous sections we have computed both the expression for the
angular bispectrum, as obtained generically in the framework of slow
roll inflation whenever one goes beyond the linear order, and the
variance associated with an unbiased estimator, assuming a mildly
non-Gaussian process. We can now compare these results for an
arbitrary configuration of $\ell_i$ multipoles. 
As a representative example, and given the fact that this was actually
the case considered in the literature, we consider 
$\ell_1 = \ell_2 = \ell_3 \equiv \ell = even$. 
We then have the bispectrum 
\be
{\cal C}_{\ell \ell \ell} =
{15\over 2\sqrt{\pi}} (3\epsilon -2\eta)
(2\ell+1)^{3/2} \; {\cal C}_{\ell}^2
\left(^{\ell~~\ell~~\ell}_{0~~0~~0}\right),
\ee
while the variance is now given by
\be
\sigma_{\hat f_{\ell\ell\ell}} = \sqrt{6} \; {\cal C}_{\ell}^{3/2}.
\ee

We show the relative amplitudes in Fig.1.  
The plot allows us to judge how plausible it is for generic one-field
inflationary models to reproduce any possible non--Gaussian structure
found on large angular scales, in particular 
on the COBE-DMR dataset. Single different values (correlations) 
for the indices $\ell_1 , \ell_2 , \ell_3$ 
(satisfying $\ell_1 + \ell_2 + \ell_3 = even$ and the triangle
inequalities) can be tried with similar result. 
Leaving aside for the time being the possibility of foreground
contamination and assuming any non--Gaussian signal is intrinsic to
the CMB, we should conclude that the presently considered class 
of models cannot explain it.

\begin{figure}
\begin{center}
\setlength{\unitlength}{1mm}
\begin{picture}(90,55)
\includegraphics{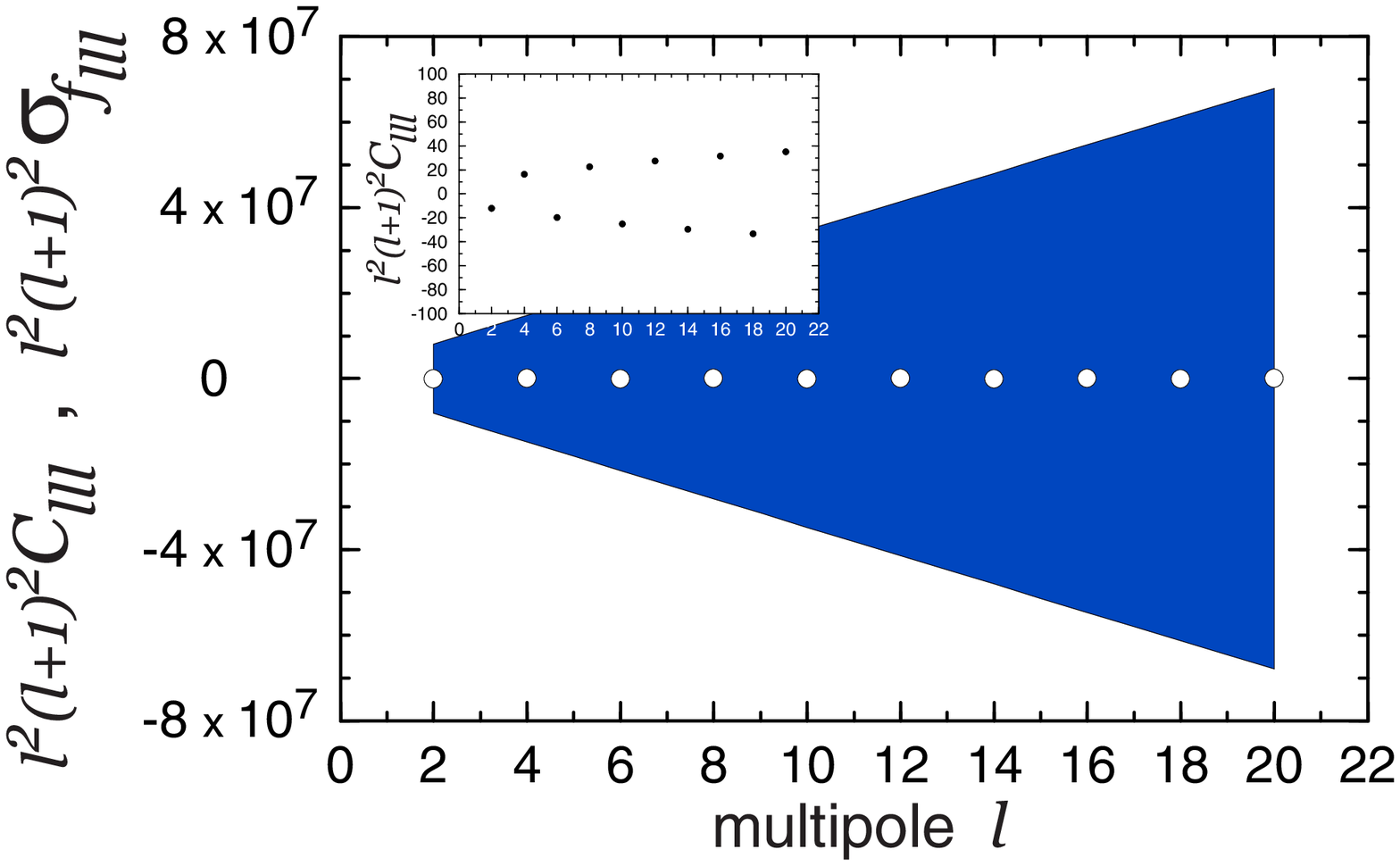}
\end{picture}
\end{center}
\vspace{1.2in}
\caption{Normalised angular bispectrum ${\cal C}_{\ell \ell \ell}$
as predicted by a generic slow roll inflation model, 
in units of ${\cal C}_{2}^2$, with ${\cal C}_{2}=1.18\times 10^{-10}$
related to the quadrupole power spectrum normalization
$Q_{\rm rms-PS} = 18.7 \mu K$, 
as a function of the multipole index $\ell$ for all even values up to
$\ell = 20$ (white dots in main plot). 
Grey band corresponds to the normalized variance 
$\sigma_{\hat f_{\ell\ell\ell}}$ (also in units of ${\cal C}_{2}^2$)
associated to the estimator of Eq. (\ref{estim}).
In the inset we zoom up $\ell^2(\ell+1)^2{\cal C}_{\ell \ell \ell}$
in the same units, which permits to see the
alternating sign of the normalised bispectrum and its actual smooth
increase in amplitude with increasing $\ell$.}
\label{figspec}
\end{figure}

\begin{figure}
\begin{center}
\leavevmode
\hbox{%
\epsfxsize=14cm
\epsffile{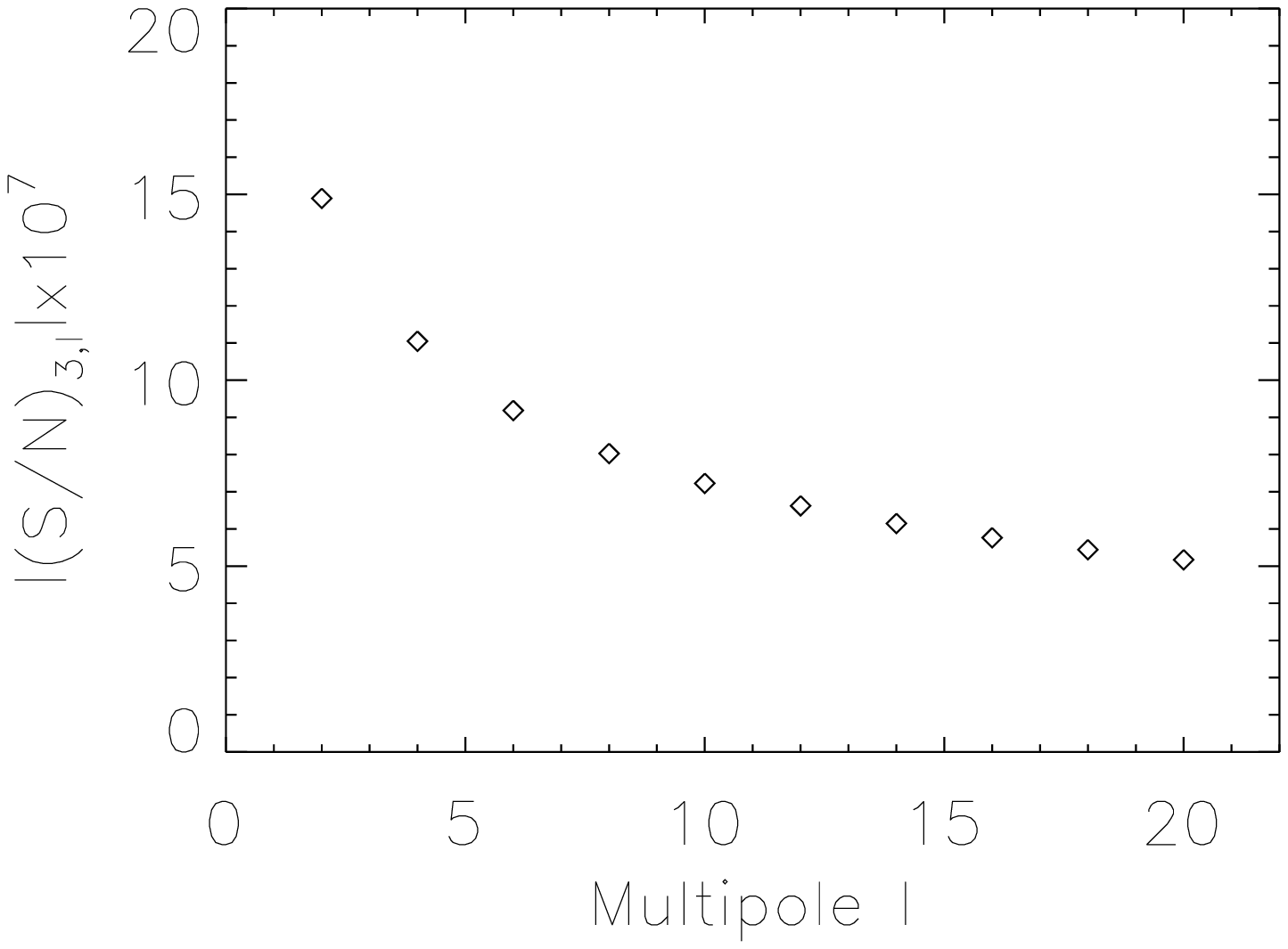}}
\end{center}
\caption{Absolute value of the signal-to-noise ratio for the bispectrum 
$(S/N)_{3,\ell}$ versus the multipole index $\ell$.}
\label{sn}
\end{figure}

\begin{figure}
\begin{center}
\leavevmode
\hbox{%
\epsfxsize=14cm
\epsffile{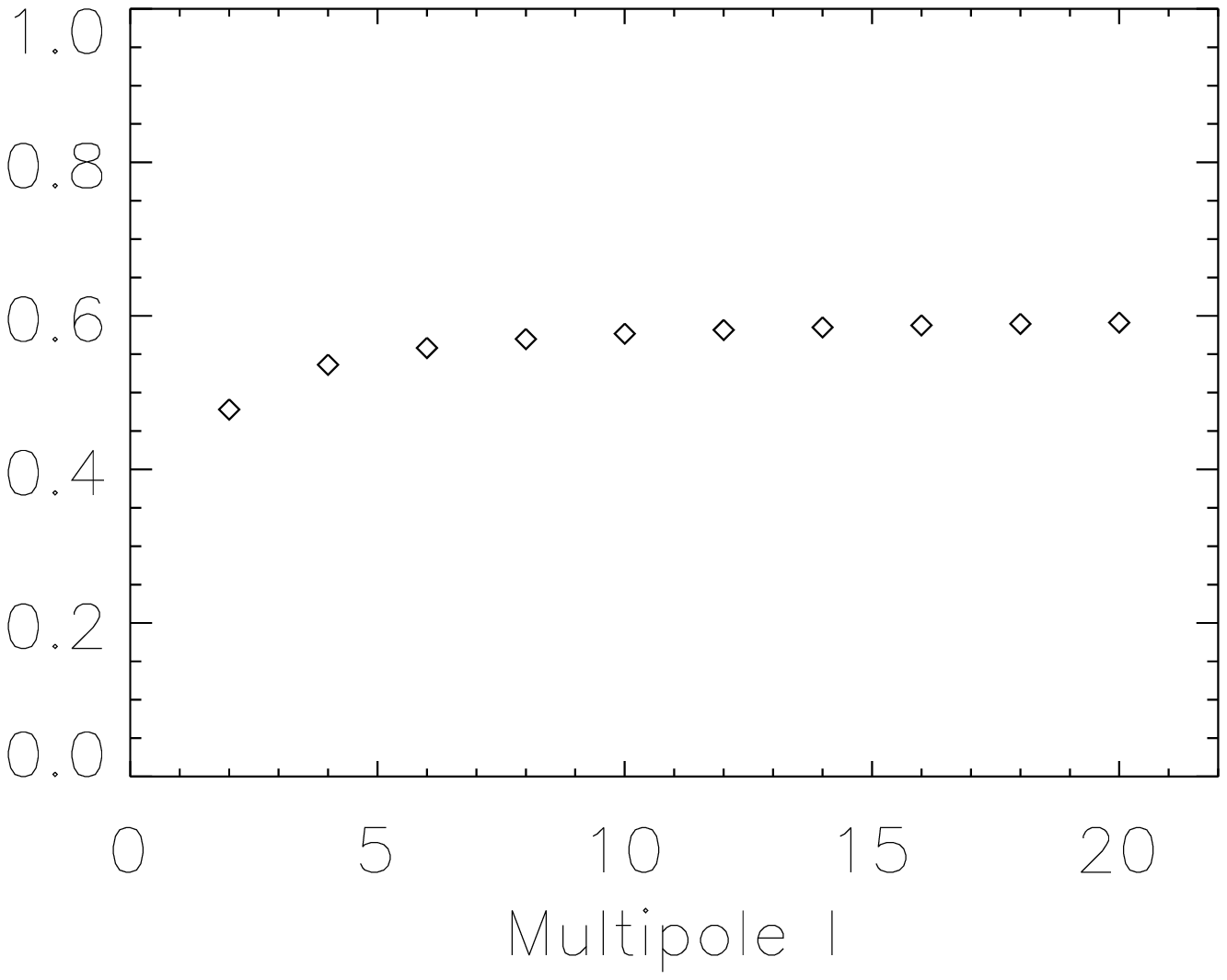}}
\end{center}
\caption{Absolute value of the product
$\ell \left(^{\ell~~\ell~~\ell}_{0~~0~~0}\right)$ for different
$\ell$'s. Note that the product vanish for all $\ell = odd$.}
\label{cleb}
\end{figure}

In order to be more specific, we now turn to the study of the
``signal-to-noise'' ratio defined by the following expressions
\be
\biggl(\frac{S}{N}\biggr)_{2,\ell}\equiv
\frac{{\cal C}_{\ell }}{\sigma_{\hat f_{\ell}}}, 
\quad 
\quad 
\quad 
\biggl(\frac{S}{N}\biggr)_{3,\ell}\equiv
\frac{{\cal C}_{\ell \ell \ell}}{\sigma_{\hat f_{\ell\ell\ell}}}, 
\ee
for the angular spectrum and bispectrum respectively. For the
first one [neglecting specificities of the particular
experiment, detector sensitivity, pixelization, etc. \cite{knox95}] 
the following well-known behaviour is found
\be
\biggl(\frac{S}{N}\biggr)_{2,\ell}=\sqrt{\frac{2\ell + 1}{2}}. 
\ee
This means that the signal emerges from the noise while going towards 
big values of $\ell$, accounting for the fact that the cosmic 
variance is important at large scales only. For the bispectrum, one has
\begin{equation}
\label{sn3}
\biggl(\frac{S}{N}\biggr)_{3,\ell}=\frac{15}{2\sqrt{6\pi }}
(3\epsilon -2\eta )(2\ell +1)^{3/2}{\cal C}_{\ell }^{1/2}
\left(^{\ell~~\ell~~\ell}_{0~~0~~0}\right) ,
\end{equation}
and we see that, contrary to the previous case,
the signal-to-noise ratio depends on ${\cal C}_{\ell }$ and  
on the slow roll parameters. The behaviour of $|(S/N)_{3,\ell}|$ is 
displayed in Fig. \ref{sn} where we see that it diminishes 
in absolute value with increasing $\ell$. This behaviour is easily
understood once we look at the hierarchical form of
${\cal C}_{\ell_1 \ell_2 \ell_3}$ and compare with the 
${\cal C}_{\ell}$ dependence of the variance. 
Again, even restricting ourselves to small $\ell$'s, we see that 
the presence of the Wigner 3-$j$ symbol is the responsible for this
particular behaviour. Hence, approaching the largest multipoles 
in the COBE-DMR data set (in particular $\ell=16$) the situation 
gets worse. In fact, from Eq. (\ref{sn3}) we roughly have
${\cal C}_{\ell } \propto \ell ^{-1}(\ell +1)^{-1}$ and
$\ell \left(^{\ell~~\ell~~\ell}_{0~~0~~0}\right)$ almost constant with
$\ell$ (see Fig. \ref{cleb}), and then 
$|(S/N)_{3,\ell}|\propto \ell ^{-1/2}$, while
$=|(S/N)_{2,\ell}|\propto \ell ^{1/2}$ in the same range of validity.
Note that the above analysis should be supplemented by a
similar one wherein the behaviour of the bispectrum coming from 
the post recombination Rees-Sciama effect is also considered. However,
given the smallness of the non-Gaussian signal this should not modify 
our conclusions very much.

To conclude, let us emphasize again the main results obtained in this
article. We have stressed that a real detection of non-Gaussianity 
in the CMB would imply that an important component of the anisotropies
arises from processes {\it other} than quantum fluctuations during an
early inflationary epoch. 
This notwithstanding, inflation predicts an actual generic form for the
bispectrum, Eq.(\ref{bisr}), 
and we here showed it explicitly in terms of the slow roll
parameters. We also computed for the first time the variance of one
candidate estimator for the bispectrum often employed in the literature 
and showed that the signal is drowned in it. Contrary to the standard
spectrum case, one cannot hope to palliate at least somewhat this 
problem by going to higher values of $\ell$ (always within the
small-$\ell$ region) since the signal-to-noise ratio decreases with 
$\ell$ like $\ell ^{-1/2}$. However, it should also be stressed that this 
conclusion might well be weaken by the finding of the actual 
best unbiased estimator. We hope to address this question elsewhere.
\par
\vspace{0.5cm}
\par
{\em Note added:} After the submission of this paper, a preprint by Wang 
and Kamionkowski ({\tt astro-ph/9907431}) appeared in which similar 
conclusions were reached.
\par
\vspace{0.5cm}
\par
\section{Acknowledgements:}

The work of A.G. was supported by the Fondation des Treilles, Paris.


\begin{thebibliography}

\bibitem[Aghanim \& Forni <1999>]{nabila}
Aghanim~N. and Forni~O., preprint {\tt astro-ph/9905124}.

\bibitem[Allen {\rm et~al.} <1997>]{alle}
Allen~B. {\rm et~al.},
Phys. Rev. Lett. {\bf 79}, 2624 (1997). 

\bibitem[Avelino {\rm et~al.} <1999>]{avel}
Avelino~P.~P. {\rm et~al}, 
Astrophys. J. Lett., preprint {\tt astro-ph/9803120}.

\bibitem[Banday {\rm et~al.} <1999>]{BZG}
Banday~A.~J., Zaroubi~S. and G\'orski~K.~M., preprint {\tt astro-ph/9908070}.

\bibitem[Bardeen, Steinhardt \& Turner <1983>]{BST}
Bardeen~J.~M., Steinhardt~P.~J. and Turner~M.~S., 
Phys. Rev. D {\bf 28}, 679 (1983).

\bibitem[ Battye {\rm et~al.} <1998>]{battye}
Battye~R.~A., Robinson~J. and Albrecht~A., 
Phys. Rev. Lett. {\bf 80}, 4847 (1998). 

\bibitem[Bersanelli {\rm et~al.} <1996>]{satellite2}
Bersanelli~M. {\rm et~al.}, Planck red book (1996), WWW page: \\
{\tt http://astro.estec.esa.nl/SA-general/Projects/Planck/ }

\bibitem[Bond \& Efstathiou <1987>]{be87} 
Bond~J.~R. and Efstathiou~G., MNRAS {\bf 226}, 655 (1987).  

\bibitem[Bouchet {\rm et~al.} <1988>]{bouchet}
Bouchet~F.~R., Bennett~D.~P. and Stebbins~A., Nature {\bf 335}, 410 (1988).

\bibitem[Bromley \& Tegmark <1999>]{bromley99} 
Bromley~B.~C. and Tegmark~M., preprint {\tt astro-ph/9904254}.

\bibitem[Bunn \& White <1997>]{Bunn97} 
Bunn~E.~F. and White~M.,  Astrophys. J {\bf 480}, 6 (1997).

\bibitem[Dodelson {\rm et~al.} <1997>]{kinney97}
Dodelson~S., Kinney~W.~H. and Kolb~E.~W., Phys. Rev. D {\bf 56}, 3207
(1997).

\bibitem[Durrer {\rm et~al.} <1996>]{dgs96} 
Durrer~R., Gangui~A. and Sakellariadou~M.,
Phys. Rev. Lett. {\bf 76}, 579 (1996).

\bibitem[Contaldi {\rm et~al.} <1999>]{conta}
Contaldi~C., Hindmarsh~M. and Magueijo~J., 
Phys. Rev. Lett. {\bf 82}, 2034 (1999).

\bibitem[Fabbri, Lucchin \& Matarrese <1987>]{flm87} 
Fabbri~R., Lucchin~F. and Matarrese~S., Astrophys. J {\bf 315}, 1 (1987).

\bibitem[Falk {\rm et~al.} <1993>]{fa93} 
Falk~T., Rangarajan~R. and Srednicki~M.,
Astrophys. J. {\bf 403}, L1 (1993).

\bibitem[Ferreira {\rm et~al.} <1998>]{ferre98} 
Ferreira~P.~G., Magueijo~J. and G{\'o}rski~K.~M., 
Astrophys. J {\bf 503}, L1 (1998).

\bibitem[Gangui <1994>]{ga94}
Gangui~A., Phys. Rev. D. {\bf 50}, 3684 (1994).

%\bibitem[Gangui, Lucchin, Matarrese \& Mollerach <1994>]{ganguietal94} 
\bibitem[Gangui {\rm et~al.} <1994>]{ganguietal94} 
Gangui~A., Lucchin~F., Matarrese~S. and Mollerach~S., Astrophys. J 
{\bf 430}, 447 (1994).

\bibitem[Gangui \& Mollerach <1996>]{gamo96} 
Gangui~A. and Mollerach~S., Phys. Rev. D. {\bf 54}, 4750 (1996).

\bibitem[Gangui \& Perivolaropoulos <1995>]{ga95} 
Gangui~A. and Perivolaropoulos~L., Astrophys. J. {\bf 447}, 1 (1995).

\bibitem[Goldberg  \& Spergel <1999>]{sperg2}
Goldberg~D. and Spergel~D., Phys.Rev. D {\bf 59}, 103002 (1999).

\bibitem[Goncharov, Linde \& Mukhanov <1987>]{GLM}
Goncharov~A.~S., Linde~A.~D. and Mukhanov~V.~F., 
Int. J. Mod. Phys. A {\bf 2}, 561 (1987). 

\bibitem[Gorski {\rm et~al.} <1996>]{Gorski96} 
Gorski~K.~M. et al, Astrophys. J. Lett. {\bf 464}, 11 (1996).

\bibitem[Grishchuk <1974>]{Gri}
Grishchuk~L.~P., Zh. Eksp. Teor. Fiz. {\bf 67}, 825 (1974) 
[Sov. Phys. JETP {\bf 40}, 409 1975)].

\bibitem[Grishchuk \& Martin <1997>]{jerogri} 
Grishchuk~L.~P. and Martin~J., Phys. Rev. D {\bf 56}, 1924 (1997).

\bibitem[Guth <1981>]{Guth}
Guth~A., Phys. Rev. D {\bf 23}, 347 (1981).

\bibitem[Guth \& Pi <1982>]{GuPi}
Guth~A. and Pi~S.~Y., Phys. Rev. Lett. {\bf 49}, 1110 (1982).

\bibitem[Hawking <1982>]{Haw}
Hawking~S.~W., Phys. Lett. B {\bf 115}, 295 (1982).

\bibitem[Heavens <1998>]{heavens98} 
Heavens~A.~F., MNRAS {\bf 299}, 805 (1998).

\bibitem[Hinshaw {\rm et~al.} <1994>]{hinshaw94}
Hinshaw~G. et al., Astrophys. J {\bf 431}, 431 (1994).

\bibitem[Hinshaw {\rm et~al.} <1995>]{hinshaw95}
Hinshaw~G. et al., Astrophys. J {\bf 446}, L67 (1995). 

\bibitem[Kamionkowski \& Kosowsky <1999>]{KK}
Kamionkowski~M. and Kosowsky~A., preprint {\tt astro-ph/9904108}.

\bibitem[Knox<1995>]{knox95} 
Knox~L., Phys. Rev. D {\bf 52}, 4307 (1995).

\bibitem[Kogut {\rm et~al.} <1996>]{kogut96}
Kogut~A. et al, Astrophys. J {\bf 464}, L29 (1996). 

\bibitem[Lange {\rm et~al.} <1999>]{balloon2}
Lange~L. {\rm et~al.}, BOOMERanG WWW page: \\
{\tt http://astro.caltech.edu/$\sim$lgg/boom/boom.html }

\bibitem[Lee {\rm et~al.} <1999>]{balloon1}
Lee~A.~T. et al, preprint astro-ph/9903249, MAXIMA WWW page: \\ 
{\tt http://cfpa.berkeley.edu/group/cmb/gen.html }

\bibitem[Liddle {\rm et~al.} <1994>]{liddle94}
Liddle~A.~R., Parsons~P. and Barrow~J.~D, Phys. Rev. D {\bf 50}, 7222
(1994).

\bibitem[Linde \& Mukhanov <1997>]{LM}
Linde~A. and Mukhanov~V., Phys. Rev. D {\bf 56}, 535 (1997).

\bibitem[Luo <1994>]{luo94} 
Luo~X., Astrophys. J {\bf 427}, L71 (1994).

\bibitem[Luo \& Schramm <1993>]{lu93}
Luo~X. and Schramm~D.N., Phys. Rev. Lett. {\bf 71}, 1124 (1993).

\bibitem[Magueijo {\rm et~al.} <1996>]{mague}
Magueijo~J., Albrecht~A., Coulson~D. and Ferreira~P.,
Phys. Rev. Lett. {\bf 76}, 2617 (1996).

%\bibitem[]{}
MAP WWW page:  {\tt http://map.gsfc.nasa.gov/ }

\bibitem[Martin \& Gangui <1999>]{Alje}
Martin~J. and Gangui~A., in preparation. 

\bibitem[Martin, Riazuelo \& Sakellariadou <1999>]{MRS}
Martin~J., Riazuelo~A. and Sakellariadou~M., 
preprint {\tt astro-ph/9904167}.

\bibitem[Mather {\rm et~al.} <1994>]{mather94} 
Mather~J.~C. et al., Astrophys. J {\bf 420}, 439 (1994). 

\bibitem[Messiah <1976>]{messiah}
Messiah~A., {\it Quantum Mechanics}, Vol. 2, Amsterdam, North-Holland.

\bibitem[Mollerach {\rm et~al.} <1995>]{molle}
Mollerach~S., Gangui~A., Lucchin~F. and Matarrese~S., 
Astrophys. J {\bf 453}, 1 (1995).

\bibitem[Munshi {\rm et~al.} <1995>]{mun} 
Munshi~D., Souradeep~T. and Starobinsky~A.~A.,
Astrophys. J. {\bf 454}, 552 (1995).

\bibitem[Pando {\rm et~al.} <1998>]{pando98}
Pando~J., Valls-Gabaud~D. and Fang~L., 
Phys. Rev. Lett. {\bf 81}, 4568 (1998).

\bibitem[Pen {\rm et~al.} <1997>]{ulpen} 
Pen~U., Seljak~U. and Turok~N., 1997, Phys. Rev. Lett. {\bf 79}, 1611.

\bibitem[Pogosian \& Vachaspati <1999>]{pova99}
Pogosian~L. and Vachaspati~T., Phys. Rev. D {\bf 60}, 083504 (1999).

\bibitem[Scaramella \& Vittorio <1991>]{sc91} 
Scaramella~R. and Vittorio~N.,  Astrophys. J. {\bf 375}, 439 (1991).

\bibitem[Smoot {\rm et~al.} <1994>]{smoot94}
Smoot~G.~F. {\rm et~al.}, Astrophys. J {\bf 437}, 1 (1994).

\bibitem[Spergel  \& Goldberg  <1999>]{sperg}
Spergel~D. and Goldberg~D., Phys.Rev. D {\bf 59}, 103001 (1999).

\bibitem[Srednicki <1993>]{sr93} 
Srednicki~M., Astrophys. J. {\bf 416}, L1 (1993).

\bibitem[Starobinsky <1982>]{Staro}
Starobinsky~A.~A., Phys. Lett. B {\bf 117}, 175 (1982).

\bibitem[Starobinski <1986>]{Starosto}
Starobinsky~A.~A, {\it Field Theory, Quantum Gravity 
and Strings}, ed. H.~J.~de Vega \& N.~Sanchez 
(Lectures Notes in Physics, Vol. 246, Berlin Springer Verlag) (1986).

\bibitem[Stewart \& Lyth <1993>]{SL}
Stewart~E.~W. and Lyth~D.~H, Phys. Lett. B {\bf 302}, 171 (1993).

\bibitem[Tegmark <1997>]{max97}
Tegmark~M., Phys. Rev. D {\bf 55}, 5895 (1997). 

\bibitem[Torres {\rm et~al.} <1995>]{torres95}
Torres~S., Cay\'{o}n~L., Mart\'{i}nez-Gonz\'{a}lez~E. 
and Sanz~J.L., MNRAS {\bf 274}, 853 (1995).

\bibitem[Turner <1993>]{turner93} 
Turner~M.~S., Phys. Rev. D {\bf 48}, 3502 (1993); ibid 5539. 

\bibitem[Unruh <1998>]{Unruh}
Unruh~W., preprint {\tt astro-ph/9802323}.

\bibitem[Verde {\rm et~al.} <1999>]{verde}
Verde~L., Wang~L., Heavens~A. and Kamionkowski~M., 
preprint {\tt astro-ph/9906301}.

\end{thebibliography}
\end{document}